\documentclass[twocolumn,showpacs,preprintnumbers,amsmath,amssymb,aps,prl,superscriptaddress,10pt]{revtex4-1}
\usepackage[latin1]{inputenc}
\usepackage{dcolumn,color}
\usepackage{bm}
\usepackage{graphicx}
\usepackage[english]{babel}
\usepackage{soul}

\usepackage{hyperref}
\hypersetup{colorlinks,linkcolor=blue,filecolor=green,urlcolor=blue,citecolor=blue}

\begin{document}

\title{Optomechanical Rydberg-atom excitation via dynamic Casimir-Polder coupling}

\author{Mauro Antezza}
\affiliation{Universit\'{e} Montpellier 2, Laboratoire Charles Coulomb UMR 5221 - F-34095 Montpellier, France}
\affiliation{CNRS, Laboratoire Charles Coulomb UMR 5221 - F-34095 Montpellier, France}
\affiliation{Institut Universitaire de France - 103, bd Saint-Michel - F-75005 Paris, France}

\author{Caterina Braggio}
\affiliation{Dipartimento di Fisica e Astronomia, Universit\`{a} degli Studi di Padova, Via Francesco Marzolo 8, I-35131 Padova, Italy}
\affiliation{INFN, Sezione di Padova, Via Francesco Marzolo 8, I-35131 Padova, Italy}

\author{Giovanni Carugno}
\affiliation{Dipartimento di Fisica e Astronomia, Universit\`{a} degli Studi di Padova, Via Francesco Marzolo 8, I-35131 Padova, Italy}
\affiliation{INFN, Sezione di Padova, Via Francesco Marzolo 8, I-35131 Padova, Italy}

\author{Antonio Noto}
\affiliation{Universit\'{e} Montpellier 2, Laboratoire Charles Coulomb UMR 5221 - F-34095 Montpellier, France}
\affiliation{CNRS, Laboratoire Charles Coulomb UMR 5221 - F-34095 Montpellier, France}
\affiliation{Dipartimento di Fisica e Chimica, Universit\`{a} degli Studi di Palermo and CNISM, Via Archirafi 36, I-90123 Palermo, Italy}

\author{Roberto Passante}\email{roberto.passante@unipa.it}
\affiliation{Dipartimento di Fisica e Chimica, Universit\`{a} degli Studi di Palermo and CNISM, Via Archirafi 36, I-90123 Palermo, Italy}
\author{Lucia Rizzuto}
\affiliation{Dipartimento di Fisica e Chimica, Universit\`{a} degli Studi di Palermo and CNISM, Via Archirafi 36, I-90123 Palermo, Italy}
\author{Giuseppe Ruoso}
\affiliation{INFN, Laboratori Nazionali di Legnaro, Viale dell'Universit\`{a} 2, I-35020 Legnaro (PD), Italy}

\author{Salvatore Spagnolo}
\affiliation{Dipartimento di Fisica e Chimica, Universit\`{a} degli Studi di Palermo and CNISM, Via Archirafi 36, I-90123 Palermo, Italy}


\begin{abstract}
We study the optomechanical coupling of a oscillating effective mirror with a Rydberg atomic gas, mediated by the dynamical atom-mirror Casimir-Polder force.
This coupling may produce a near-field resonant atomic excitation whose probability scales as $\propto (d^2\;a\;n^4\;t)^2/z_0^8$, where $z_0$ is the average atom-surface distance, $d$ the atomic dipole moment, $a$ the mirror's effective oscillation amplitude, $n$ the initial principal quantum number, and $t$ the time.  We propose an experimental configuration to realize this system with a cold atom gas trapped at a distance $\sim 2\cdot10 \, \mu$m from a semiconductor substrate,  whose dielectric constant is periodically driven by an external laser pulse, hence  realizing en effective mechanical mirror motion due to the periodic change of the substrate from transparent to reflecting. For a parabolic gas shape, this effect  is predicted to excite  about $\sim 10^2$  atoms of a dilute gas of $10^3$ trapped Rydberg atoms with $n=75$  after about $0.5 \,\mu \mbox{s}$, hence high enough to be detected in typical Rydberg gas experimental conditions.
\end{abstract}

\maketitle

\emph{Introduction.}
Zero-point fluctuations are among the most striking consequences of the quantum description of the electromagnetic field. They are at the origin of the Casimir-Lifshitz force, that is a long-range quantum electromagnetic interaction between neutral polarizable bodies \cite{Milonni,CPP,Casimir}. This interaction is relevant for both fundamental and applicative purposes, and has been extensively investigated both theoretically and experimentally for several configurations  (see \cite{Casimir2} and references therein). For the atom-surface configuration, it takes the name of Casimir-Polder (CP) force, and at $T=0$ several regimes are present: a very short-distance regime, for distances comparable to the surface plasma wavelength, where the dielectric properties of a real surface can be important, a near-zone regime  (van der Waals or nonretarded Casimir-Polder) for atom-wall distances smaller than a typical transition wavelength of the atom, and a long-distance regime (retarded Casimir-Polder) for larger distances. For typical short wavelength atomic transitions, there is also a thermal regime dominating at large separations. For long wavelength transitions (molecular or Rydberg states, as in this paper) and for the conditions considered in this paper, this thermal regime is absent \cite{Thermal}.
When boundary conditions are set in motion with nonuniform acceleration in the vacuum, or when material properties are changed nonadiabatically, a dynamical Casimir effect is realized, and a parametric excitation of vacuum fluctuations may lead to the emission of real photons \cite{DCE}. Similarly, a dynamical Casimir-Polder effect occurs when physical parameters of an atom near a conducting plate are rapidly changed \cite{halfdressed}.
The dynamical Casimir effect and its analogues are delicate effects due to the small number of emitted photons, and have been recently observed in superconducting circuits \cite{WJP11}, in Josephson metamaterials \cite{LPHH13} and in Bose-Einstein condensates \cite{JPB12}.

Another rapidly growing research field is that of quantum optomechanics, which deals with systems where mechanical degrees of freedom are coupled to cavity fields \cite{Meystre13}. Such systems have been experimentally and theoretically investigated, for example, for realizing sensitive force detectors, cooling macroscopic mirrors or obtaining quantum superposition states for macroscopic objects \cite{AKM13}. Significant experimental progresses have been obtained in precision trapping of cold atoms near a nanoscale optical cavity, allowing to probe cavity near fields \cite{Thompson13}. The effect of quantum fluctuations of the position of a cavity mirror on Casimir and Casimir-Polder interactions has been also demonstrated \cite{BP13}.

In this Letter we propose a new optomechanical Rydberg atoms-surface coupling based on a novel aspect of the dynamical CP effect, able to affect the internal atomic state. It is a near field effect, not related to the excitation of atoms by the few real photons expected in the dynamical Casimir effect \cite{Dynamical1,Dynamical2}. We consider a gas of dilute Rydberg atoms  trapped in front of a substrate whose refractive index is changed in time (dynamical mirror) at a frequency corresponding to one of their transition frequencies. Due to the effective periodical change of the atom-mirror distance, the optomechanical coupling between the wall and the Rydberg atoms yields a periodic perturbation on the atoms, which can be excited to upper levels.
On the experimental side,  this scheme may largely profit from recent progresses in the realization of dynamical mirrors \cite{ABCVGMPRR11}, and in the cigar-shape trapping of Rydberg atoms and their preparation in long-lived excited states \cite{RYCLKBR13}.

It is worth saying that recently a micromechanical atom-wall system has been realized with a trapped Rb BEC close to a dielectric substrate, and the collective oscillations of the gas have been used to measure the CP force \cite{APS04,Antezza06}. In particular this allowed the first measurement of the more elusive thermal component of this interaction \cite{APS05,OWAPSC07}. Differently from that case, where the \emph{external} degrees of freedom of an atomic gas have been used \emph{to detect} the CP force, here we \emph{use} the CP force to couple a substrate to the \emph{internal} atomic degrees of freedom.

\emph{The Model.}
We consider a fixed Rydberg atom near a perfectly conducting plate; the plate is forced to move harmonically around its equilibrium position. We model the atom as a two-level system. The mirror's position coincides with the plane $z = 0$ at $t = 0$, and the atom-mirror distance is $z(t)$. We first analyze the case of a fixed mirror at a distance $z$ from the atom; we assume that this distance is much smaller than a main transition wavelength $\lambda_0=2\pi c/\omega_0$ of the Rydberg atom, $\omega_0$ being the corresponding transition angular frequency. The atom-mirror CP interaction energy is thus in its near-zone nonretarded regime, where electrostatic (longitudinal field) contributions are dominant \cite{Milonni,APS04}. We assume the atom prepared in a long-lived Rydberg state and treat it as a stable state, assuming to study the system for times shorter than its lifetime. The CP interaction energy between a ground-state atom and a fixed perfectly conducting mirror, within dipole approximation and in the nonretarded regime, is \cite{Buhmann12,BAM00}
\begin{equation}
V(z) =-\frac {\langle d_x^2 \rangle +\langle d_y^2 \rangle + 2\langle d_z^2 \rangle }{16 z^3}= -\frac{1}{16}\frac{\sigma_{ij}\langle d_i d_j \rangle }{z^3},
\label{eq:1}
\end{equation}
where sum over repeated indices is used together with {\it cgs} units, and the average of the squared components of the atomic dipole moment operator ${\bf d}$ are taken on the atomic state considered and the atom-mirror distance $z$ is along the $\hat{z}$ direction. We have also defined the diagonal matrix $\sigma = \mbox{diag}(1,1,2)$.

The expression \eqref{eq:1} of the atom-wall interaction for an ideal conductor is a very good approximation in our case, since we shall consider atom-wall distances of the order of $2 \cdot 10^{-3} \, \mbox{cm}$, much larger than the plasma wavelength of a typical metal (of the order of
$\lambda \sim 10^{-5} \, \mbox{cm}$), where real-conductor corrections are known to be negligible \cite{IHA11}.

It is well known that in the near-zone limit, the atom-wall nonretarded interaction \eqref{eq:1} is well described by the interaction between the atomic dipole and its image  \cite{BAM00}.
In order to describe the interaction of the atom with the oscillating mirror, we adopt a semiclassical model: we obtain the atom-wall interaction as the interaction energy between the atomic dipole and an effective classical field due to the image atom. Using the method of image charges \cite{Jac75}, we can describe the near-zone atom-wall interaction by the coupling term
$H_I=-\mathbf{d}\cdot{\mathbf{E(\mathbf{r})}}/2$,
where $\mathbf{d}$ is the atomic dipole moment operator and $\mathbf{E(\mathbf{r})}$
is the electric field generated by the image dipole $\tilde{d} =(-d_x, -d_y, d_z)$ at the atom's position $\mathbf{r}=(0,0,z)$. The factor $1/2$ takes into account that, when the atomic dipole is moved from infinity to its final position, the image dipole moves symmetrically in the opposite direction \cite{PT82b}. Comparing $H_I$ with
Eq. \eqref{eq:1}, the effective electric field acting on the atom, due to the image atom at a distance $2z$, is
\begin{equation}
\label{eq:6}
E_i(\mathbf{r})=\frac{1}{8}\frac{\sigma_{ij}d_j}{z^3} \, .
\end{equation}
Quantum mechanically, a fluctuating field is indeed acting on the atom,
due to vacuum fluctuations modified by the presence of the conducting wall. Its effect on the atom is equivalent to the action of the effective field \eqref{eq:6}.
The effective semiclassical interaction Hamiltonian, quadratic in the atomic dipole moment operator, is then
\begin{equation}
\label{eq.5}
H_I= - \left( \frac {\sigma_{ij}}{16z^3}\right)d_i d_j \, ,
\end{equation}
whose average on the atomic state coincides with \eqref{eq:1}.

We now assume that the plate oscillates harmonically around its equilibrium position $z=0$, so that the atom-wall distance changes as $z(t)=z_0[1-\frac{a}{z_0}\sin\omega t]$, where $z_0$ is the average atom-mirror distance and $\omega$ the mirror's oscillation angular frequency. We also assume the oscillation amplitude $a$ such that $a\ll z_0$, and that the atom-wall Casimir-Polder interaction follows instantaneously the plate oscillation. The latter assumption is fully consistent with our near-zone hypothesis, because the interaction is non-retarded in this case. From \eqref{eq.5}, the effective interaction Hamiltonian between the atom and the oscillating mirror will then take the form
$H_I(t)=H_s+V_I(t)$, where $H_s$ is a time-independent term yielding the usual $z_0^{-3}$ near-zone static Casimir-Polder potential \eqref{eq:1}, and
\begin{equation}
\label{eq:7}
V_I(t)\simeq-d_id_j\left(\frac {3\sigma_{ij}}{16z_0^3}\frac a{z_0}\sin\omega t\right)\, ,
\end{equation}
is a time-dependent term linked to the plate oscillation. The approximation $\frac 1{z^3(t)}\simeq\frac 1{z_0^3}(1+\frac{3a}{z_0}\sin\omega t)$ has been used.
The perturbation \eqref{eq:7} can induce transitions in the atom.
Applying time-dependent perturbation theory with $V_I(t)$ as the perturbation operator, we can easily obtain the probability amplitude for the transition  $|g\rangle\rightarrow|e\rangle$ from the atomic initial state $g$ to a more excited state $e$,
\begin{equation}
\label{eq:8}
c_e(z_0,t)
=\frac i\hbar \frac{3\sigma_{ij} (d_i d_j)^{eg}}{16z_0^3}\frac{a}{z_0}\int_0^{t}dt'e^{i\omega_0t'}\sin\omega t' \, ,
\end{equation}
where $\hbar \omega_0 = E_e - E_g$ and $(d_i d_j)^{eg}=\langle e \mid d_i d_j \mid g \rangle$. Under resonance conditions ($\omega\simeq\omega_0$), the atomic excitation probability $P_e(z_0,t)=|c_e(z_0,t)|^2$ is given by
\begin{equation}
\label{eq:9}
P_e(z_0,t)\simeq\frac{9}{2^{10}\hbar^2}\left(\frac{a}{z_0}\right)^2\frac{\mid \sigma_{ij} (d_i d_j)^{eg} \mid^2}{z_0^6}t^2.
\end{equation}

We now estimate the order of magnitude of the excitation probability $P_e(t)$. For a Rydberg atom, the matrix element of the product of components of the atomic dipole moment appearing in \eqref{eq:9} is related to the electron charge $e$, the Bohr radius $a_0$ and the principal quantum number $n$ through the relation
$\mid \sigma_{ij}d_i d_j \mid \sim e^2 a_0^2 n^4$ \cite{SECSBS10,Gallagher88}. Thus the excitation probability approximately becomes

\begin{equation}
\label{eq:10}
P_e(z_0,t) \simeq \left( 3 \cdot 10^{-19} \mbox{cm}^6\mbox{s}^{-2} \right) \frac{a^2}{z_0^8} n^8 t^2.
\end{equation}

The condition $P_e(z_0,t)\ll 1$ must be satisfied for the validity of our perturbative approach, and this sets an upper limit to the acceptable values of time $t$, once the other parameters have been fixed.
For $n=75$, yielding a frequency of about $30$ GHz for the transition $n=75 \rightarrow 77$,  $a/z_0 \simeq10^{-1}$ and $z_0 \simeq 2 \cdot 10^{-3} \, \mbox{cm}$, single atom excitation probability $
P_e(t) \simeq  \left(5\cdot 10^{10} \, \mbox{s}^{-2}\right) t^2$ shows that, by taking a time of the order of $2$\,$\mu$s (well compatible with achievable Rydberg atoms trapping times \cite{VBRMCMA11,Pillet09}), the probability is of the order of $20 \%$.

If we consider now a trapped Rydberg gas instead of a single atom, if the trap size is comparable with the atom-mirror distance Eq. \eqref{eq:10} could not be enough accurate, and the actual profile of the atomic trap should be taken into account.
If $\rho(z)$ is the atomic linear density in the direction $z$ orthogonal to the surface, the number of excited atoms, neglecting interactions among them, can be written as
\begin{equation}
\label{eq:12}
N_e(t)=\int_0^\infty dz\rho(z)P_e(z,t) \, .
\end{equation}
If the gas profile is cigar-shaped parallel to the mirror, as a first approximation we may use a parabolic profile in the three dimensions.
Let $N$ be the number of Rydberg atoms in the gas, $z_c$ and $2R_z$ (with $R_z < z_c$) respectively the trap center-wall distance and the width of the gas profile along $z$, the atomic linear density in the $z$ direction is given by
\begin{equation}
\label{eq:12a}
\rho(z) = \frac{3N}{4R_z^3} \left[ R_z^2 - (z-z_c)^2\right]
\end{equation}
for $-R_z < z - z_c <R_z$, and $\rho (z) = 0$ for $z-z_c < -R_z$ or $z-z_c > R_z$. Then from \eqref{eq:10}, \eqref{eq:12} and \eqref{eq:12a}, the number of excited atoms at time $t$ is
\begin{eqnarray}
\label{eq:12b}
N_e(t) &\simeq& \left( 10^{-20} \mbox{cm}^6 \, \mbox{s}^{-2} \right)
\nonumber \\
&\times& \frac {3 +42 \bar{z}_c^2 +35\bar{z}_c^4}{\left( \bar{z}_c^2-1 \right)^6} \frac{N a^2 n^8 t^2}{R_z^{8}} ,
\end{eqnarray}
where $\bar{z}_c=z_c/R_z$.
Other gas profiles (gaussian, for example) could be directly used in \eqref{eq:12} in order to refine our estimate to specific experimental setups.

\begin{figure}[h!]
\begin{center}
\includegraphics[width=2.5in]{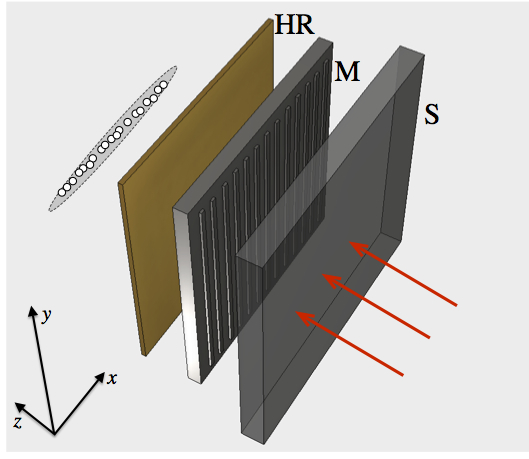}
\caption{(Color online) Exploded scheme of the system:
The trapped Rydberg atoms interact with the dynamical mirror (M), made by a semiconductor layer whose rear surface is covered by a metallic mesh. The semiconductor layer is illuminated by a multigigahertz repetition rate laser that induces a periodical variation of its dielectric properties. This dynamical mirror is sandwiched between a transparent bulk dielectric (S) that acts as thermal sink, and a Bragg high reflectivity mirror (HR).}
\label{ufa}
\end{center}
\end{figure}

\emph{Experimental proposal.}
In Fig.\,\ref{ufa} a possible experimental scheme to realize the proposed optomechanical coupling is shown.
The required dynamic mirror is based on a semiconductor layer whose conductivity is changed by a train of laser pulses \cite{Padova09} impinging on its rear surface, covered by a metal mesh that acts as a \emph{rear} mirror.
On the opposite side of the semiconductor, a high reflectivity interferential dielectric Bragg mirror, tuned to the the incident laser wavelength, prevents light that has not been absorbed in the semiconductor to enter the Rydberg atoms region.
The thickness of the rear metal mesh should assure a good reflectivity at all field frequencies relevant for the Casimir-Polder interaction at the considered atom-mirror distance $z_0$.
In the figure a parallel lines structure mesh is shown, even if more complex patterns can be designed.
To optimize the dynamical mirror, the size of the non-metallic areas in the mesh could be further reduced well below the incident laser wavelength. The laser beam would then be transmitted through the rear mirror by exploiting the EOT (extraordinary optical transmission) effect \cite{Garcia10}.
In the devised scheme $a$ is comparable with the thickness of the semiconductor layer, of the order of a few micrometers, which is in turn comparable with $\alpha^{-1}$, the absorption length of near infrared light in direct band gap semiconductors. For example, $\alpha^{-1}\sim 1$\,$\mu$m in GaAs excited at the band gap photon energy that corresponds to $\lambda\sim800$\,nm.

The Rydberg atoms are prepared in an initial state characterized by a principal quantum number $n$, which determines the required oscillation frequency of the mirror in order to obtain a resonance effect for the transition to an higher energy level.
We could assume that the initial state of the Rydberg atom is a circular state (maximum angular momentum quantum numbers), yielding a very long lifetime of the initial state. Angular momentum selection rules for the transition give $\Delta \ell = 0, \pm 2$. A possible transition worth to consider in our case is that with $\Delta \ell = 2$. By tuning the mirror frequency we can set also $\Delta n = 2$. Such a transition brings the Rydberg atom to a final state with maximum azimuthal quantum number, that is a long-lived state too, because only one decay channel by a dipole transition is allowed \cite{Gallagher88}. This should make easier the detection of the atomic excitation.
If, as mentioned, an initial $n=75$ state is prepared, the atom should be promoted to the upper level $n=77$ when the moving mirror oscillates at a frequency of approximately 30 GHz. A MOPA (master oscillator power amplifier) laser system \cite{ABC08} with a seed oscillator operating at 30\,GHz delivers the pulses at the required repetition rate, and with an energy/pulse (few $\mu$J ) sufficient to excite a plasma mirror in the semiconductor layer.
Lower mirror oscillation frequencies (thus lower repetition rates) would be allowed for initial states with a higher principal number. A limitation to $n$ is however set by the detection technique of the excited states, which would rely on the selective field ionization \cite{Gallagher88}.
In previous experiments using Rydberg states with $n=30-85$ a field control at the mV/cm level has been demonstrated \cite{Vit11}.
On the other hand, the requirements for the wall vibration frequency become ever more stringent for decreasing initial quantum numbers.

The non-harmonicity of the atom-wall distance in the presented experimental scheme can be easily included by using the atom-mirror distance $z(t)=z_0[1-\frac{a}{z_0}f(t)]$, where $f(t)$ is the appropriate function describing the mirror's motion. In this case, Eq. \eqref{eq:8} becomes
\begin{eqnarray}
\label{eq:11}
c_e(z_0,t)
&=& -\frac{1}{\sqrt{2\pi}\hbar} \frac{3 \sigma_{ij}(d_i d_j)^{eg}}{16z_0^3}\frac{a}{z_0} \nonumber\\
&\ &\times\int_{-\infty}^{\infty}d\omega g(\omega)\frac{e^{-i(\omega-\omega_0) t}-1}{\omega-\omega_0},
\end{eqnarray}
where $g(\omega)$ is the Fourier transform of $f(t)$. Once its form is obtained for the specific experimental setup considered, the squared modulus of Eq. \eqref{eq:11} gives the corresponding atomic excitation probability, generalizing Equation \eqref{eq:9}.

In our experimental proposal, the optomechanical coupling with the moving surface could be optimized with a quasi-one-dimensional Rydberg gas prepared in a magneto-optic trap. Such a Rydberg gas has been recently obtained in a trap with a density of $10^{10}$ atoms/cm$^3$ and number of atoms reaching the order of $10^6$,  whose distance from a surface could be controlled with a few micrometers precision \cite{RYCLKBR13,VBRMCMA11}. Lower densities are preferable in our case, in order to reduce possible interactions among the atoms.
The average atom-wall distance we consider, in the range $20-50$ micrometers, is much less than a typical transition wavelength of the Rydberg atom (of the order of $1$ cm for the transition mentioned above), so that the near-zone Casimir-Polder interaction between the atoms and the mirror is relevant in this case, as we have assumed.

Trapping a sample of about $N \sim 10^3$ $^{87}$Rb atoms in a cigar-like shape with $R_{\perp}\sim 5\cdot 10^{-2}\, \mbox{cm}$ and $R_z \sim 10^{-3}\, \mbox{cm}$ at a distance $z_c=2 \cdot 10^{-3}\, \mbox{cm}$ from a surface for times up to $10$\,$\mu$s, is also realistic using actual Rydberg atoms trapping techniques.

From Eq.\eqref{eq:12b}, with the numbers above and for $a=  2 \cdot 10^{-4} \mbox{cm}$, we obtain that about $100$ atoms in the sample of $10^3$ are excited after $0.5 \, \mu \mbox{s}$ (atoms in the sample closer to the wall are more likely excited, due to the strong dependence of the excitation probability with the atom-wall distance). Hence a considerable number of trapped atoms can be excited in a quite short time interval.

The atoms density considered in the previous estimate is such that the long-range atom-atom van der Waals interaction, which scales as $r^{-6}$ with the interatomic distance $r$, can be neglected. In fact, the closest atom-atom distance is around $10^{-3}\, \mbox{cm}$ and the average atom-wall distance is $2 \cdot 10^{-3}\, \mbox{cm}$. Using known expressions  \cite{quadrupolar} it is possible to show that the atom-wall interaction energy is some three orders of magnitude larger than the interaction energy between one atom and its closest atom.
Similarly, it is easy to check that quadupolar interactions \cite{quadrupolar,HFZ86} are several orders of magnitude smaller than dipolar ones, and can be therefore neglected. This is also expected on a physical basis, because in our case the size of the Rydberg atoms, $\sim 10^{-5} \, \mbox{cm}$, is much smaller than the average relevant atom-wall and atom-atom distances.

Finally, we can compare our excitation probability of the Rydberg atoms with that due to absorption of the real photons emitted by dynamical Casimir effect. Using known results for the number of emitted photons by an oscillating wall \cite{photon}, with the same parameters given above for our proposed experiment, the number density (for unit area) of real photons emitted is $\sim 10^{-1}/ \text{cm}^2$.
Then the number of photons that can excite our trapped atomic sample, which has a front area of $\sim 10^{-5} \text{cm}^2$, is $\sim 10^{-6}$. This is an upper limit for the single-atom excitation probability by the real photons (far field) emitted by the oscillating wall under resonance condition, and it is thus negligible compared to our near-field excitation probability $(\sim 10 \%)$. The number of emitted photons could be increased by a resonant cavity \cite{Dynamical1}, but also in this case the atomic excitation probability is order of magnitudes smaller than our near-field effect.

\emph{Conclusions.}
We propose a new optomechanical dynamical coupling between Rydberg atoms and a substrate, based on the dynamical Casimir-Polder effect. In particular, we have analyzed a dilute sample of Rydberg atoms trapped in the proximity (tens of micrometers) of an oscillating reflective mirror. This effect could be observed using currently available experimental techniques.

The mirror's mechanical oscillation can be conveniently simulated by a semiconductor slab with a time-varying dielectric constant (dynamical mirror, periodically switching from dielectric to conductor), obtained by appropriate laser pulses. Due to the mirror's effective oscillation, the atoms are subjected to a time-dependent nonretarded Casimir-Polder interaction potential (optomechanical coupling), which can induce transitions of the atoms to an upper level following an analytical power law we derived in a semiclassical approximation and for a given gas profile. The atomic excitation probability results to be significant and detectable in typical experimental conditions, that we have discussed in detail. This effect shows how quantum vacuum fluctuations may be used to realize an optomechanical coupling between a macroscopic body and an elementary or mesoscopic quantum system, and to change its internal state.

\emph{Acknowledgments.}
The authors acknowledge F. Capasso, V. Dodonov, B. Guizal, C. Henkel, R. Messina, P. Pillet and M. Weidem\"{u}ller for stimulating discussions and suggestions on the subject of this paper.
Financial support by the Julian Schwinger Foundation, by MIUR and by CRRNSM is gratefully acknowledged.

\end{document}